\documentclass[10pt,conference]{IEEEtran} 

\usepackage[utf8]{inputenc}
\usepackage[T1]{fontenc}
\usepackage{braket}
\usepackage{booktabs}
\usepackage{amsmath}
\usepackage{amssymb}
\usepackage{graphicx}
\usepackage{algorithm}
\usepackage{algorithmic}
\usepackage{pgfplots}
\pgfplotsset{compat=1.18}
\usepgfplotslibrary{groupplots}
\usepackage{geometry}
\usepackage{todonotes} 

\title{Quantum Counterparty Credit Risk: A Study of Path-Dependent Derivatives}
\author{
    \IEEEauthorblockN{
        Sandeep Jha\IEEEauthorrefmark{1}, 
        Richard Oentaryo\IEEEauthorrefmark{1},
        Sanjay Sekaran\IEEEauthorrefmark{1}, 
        Vadym Kullish\IEEEauthorrefmark{1},
        Rajanikanth Annam\IEEEauthorrefmark{1},\\
        Paul Griffin\IEEEauthorrefmark{2},
        Hanan Rosemarin\IEEEauthorrefmark{3},
        Nadav Yoran\IEEEauthorrefmark{3}, 
        Nati Erez\IEEEauthorrefmark{3}, 
        Rei Sato\IEEEauthorrefmark{4}
    }
    \IEEEauthorblockA{
        \IEEEauthorrefmark{1}DBS Bank Ltd., Singapore \enspace
        \IEEEauthorrefmark{2}Singapore Management University, Singapore \\
        \IEEEauthorrefmark{3}Classiq Technologies Ltd., Tel Aviv, Israel \enspace
        \IEEEauthorrefmark{4}Classiq Technologies G.K., Tokyo, Japan        
    }
    \thanks{Corresponding author: Sandeep Jha (sandeepjha@dbs.com).}
}
\date{\today}

\begin{document}

\maketitle

\begin{abstract}
Estimating potential future exposure (PFE) for path-dependent derivatives, such as FX Target Redemption Forwards (TARFs), represents a formidable computational challenge due to the demand of nested Monte Carlo simulations. We present a hybrid quantum--classical framework that leverages Iterative Quantum Amplitude Estimation (IQAE) to address this via a reduced-order counterparty credit risk model. Our methodology maps the non-linear TARF payoff---including cumulative gains and knock-out features---into a quantum circuit via a two-step formulation, whereby a first-step percentile is computed classically and then used to condition quantum evaluation of subsequent exposure. We employ discretisation of the FX process and a linearised additive approximation of dynamics to enable implementation on current quantum platforms. Developed via the Classiq platform and validated on NVIDIA CUDA-Q and Amazon Braket SV1, our approach achieves relative errors of $1\%$–-$8\%$ against classical benchmarks at the $97.5\%$ and $99\%$ confidence levels. While discretisation constraints and approximate monotonicity assumption may introduce bias and limit recovery of the full exposure distribution, our framework offers a tractable testbed for quantum acceleration. Scaling analysis suggests that \textasciitilde{300} logical qubits could enable full 52-week exposure estimation, reducing sample complexity for tail-risk estimation via amplitude estimation at the cost of increased circuit depth.
\end{abstract}

\section{Introduction}

The resilience of the global financial system hinges on high-fidelity quantification of risk within the derivatives market \cite{leRoux2008}. Unlike traditional loans with deterministic exposure profiles, derivative portfolios represent stochastic processes that fluctuate with underlying market volatility. Under the Basel III/IV frameworks, \textit{Counterparty Credit Risk (CCR)} is a key determinant of bank capital. Financial institutions operating in the international monetary market are mandated to estimate complex risk metrics, such as \textit{Expected Exposure (EE)} and high-quantile \textit{Potential Future Exposure (PFE)}, across multi-year horizons. Given the scale of current derivative exposures, even marginal inaccuracies in these revaluations yield a substantial amount of regulatory capital and limit-utilisation impact \cite{Ghamami2014}.

The most acute computational bottleneck lies in path-dependent exotics, such as the \textit{FX Target Redemption Forward (TARF)}. Valuing these instruments is an inherently non-linear challenge, as the payoff is governed by the cumulative temporal history of exchange rates and discrete "knock-out" triggers that introduce sharp discontinuities in the state space. Traditionally, estimating PFE for such instruments typically requires \textit{nested Monte Carlo} (MC) simulations to evaluate conditional expectations across future states. Despite the maturity of parallel CPU/GPU architectures, these methods remain bound by sub-optimal convergence rates---often scaling as $\mathcal{O}(1/{N^3})$ in nested configurations---creating a systemic ceiling for real-time risk sensitivity analysis \cite{AbbasTurki2018, glasserman2004monte}.

Quantum computing offers an algorithmic departure from this complexity class. \textit{Quantum Amplitude Estimation (QAE)} provides a provable quadratic speed-up, improving error convergence to $\mathcal{O}(1/N)$ \cite{brassard2002quantum, montanaro2015quantum, woerner2019quantum}. However, the \textit{unitary encoding} of this problem remains a challenge; translating the discontinuous triggers and path-dependent accumulation of an FX TARF into a coherent quantum state requires a sophisticated mapping of classical "if-then" branching onto quantum operators---a task that pushes the limits of modern circuit synthesis. Furthermore, realizing this quantum utility is constrained by the \textit{Noisy Intermediate-Scale Quantum (NISQ)}-era hardware, and requires a strategic balance between problem discretization and the constraints of gate fidelity and qubit count.

This work introduces a hybrid quantum–classical framework implemented on a reduced two-step exposure model---a deliberate scoping choice that preserves the essential path-dependent and knock-out characteristics of the FX TARF while ensuring tractability on current infrastructure. Rather than attempting to perform full distributional estimation of exposure, we selectively apply quantum subroutines to the conditional expectation component of the CCR workflow. This validated framework serves as a basis for projecting full-horizon resource requirements. We summarize our contributions below:

\begin{itemize}
    \item \textbf{Quantum Encoding of Path-Dependency:} We develop a circuit architecture encoding the non-linear ``knock-out'' logic of FX TARFs into a unitary operator, providing a scalable template extensible to longer horizons.
    \item \textbf{High-Fidelity Validation:} We show that Iterative QAE (IQAE) recovers PFE at the 97.5\% and 99\% confidence levels with 1-8\% relative error against classical MC benchmarks, confirming that the quantum estimation procedure itself is not the dominant source of inaccuracy.
    \item \textbf{Heterogeneous Emulation Benchmarking:} We analyze trade-offs between circuit depth and estimation precision across \textit{NVIDIA CUDA-Q} and \textit{Amazon Braket}, as discretization resolution scales from 4 to 6 qubits.
    \item \textbf{Quantum Advantage Projection:} Grounded in the empirical scaling of the two-step model, we project that a fault-tolerant quantum system with $\sim$300 logical qubits would enable full 52-week horizon estimation, rendering IQAE meaningfully advantageous over nested MC.
\end{itemize}

\section{Related Work}

The landscape of quantitative finance is undergoing a major shift as the limitations of classical stochastic modeling meet the theoretical potential of quantum information science. This section reviews the evolution of counterparty risk modeling and the parallel development of quantum algorithms for accelerating these high-dimensional computations.

The foundational framework for MC methods in financial engineering was laid out by Glasserman~\cite{glasserman2004monte}, establishing MC simulation as the industry standard for pricing derivatives. In the domain of CCR, Ghamami and Zhang~\cite{Ghamami2014} provided requisite frameworks for Credit Valuation Adjustment~(CVA) and exposure measurement widely adopted by global institutions today. However, the $\mathcal{O}(1/\sqrt{N})$ convergence rate of classical MC becomes very costly in the \textit{nested} setting, where derivative payoffs must be evaluated at every future simulated time step~\cite{AbbasTurki2018}. For path-dependent instruments such as the FX TARF, this complexity is further compounded by knock-out features and cumulative payoff logic. While the industry-standard Least Squares Monte Carlo~(LSMC) method~\cite{Longstaff2001} avoids nesting by regressing continuation values, our approach addresses the same bottleneck---conditional expectation estimation---via amplitude estimation. This provides an alternative that becomes beneficial at high quantile levels where rare-event sampling typically dominates LSMC's error budget.

The theoretical foundation for quantum speedup in statistical sampling was established by Brassard et al.~\cite{brassard2002quantum} through QAE, which reduces sample complexity from $\mathcal{O}(\epsilon^{-2})$ to $\mathcal{O}(\epsilon^{-1})$. Montanaro~\cite{montanaro2015quantum} later formalized this speedup for general MC methods, proving that quantum computers can in principle outperform classical samplers in high-dimensional integration. To bridge the gap between theory and NISQ-era constraints, research has shifted towards QAE variants that eliminate Quantum Phase Estimation~(QPE). The most notable examples include Iterative QAE~(IQAE) and Maximum Likelihood Amplitude Estimation~(MLAE), which exhibit substantially reduced circuit depth~\cite{grinko2021iterative, suzuki2020amplitude}.

The first application of QAE to financial risk management was presented by Woerner and Egger~\cite{woerner2019quantum}, who
demonstrated evaluation of Value-at-Risk~(VaR) and Conditional Value-at-Risk~(CVaR) on gate-based quantum processors by encoding log-normal return distributions into quantum states~\cite{stamatopoulos2020option, Zoufal2019}. In credit-specific contexts, Egger et al.~\cite{Egger2020} extended these concepts to Credit Risk Analysis~(CRA), while Alcazar et al.~\cite{Alcazar2022} introduced CVA estimation workflows via quantum subroutines. Stamatopoulos et al.~\cite{Stamatopoulos2022} further identified market sensitivity~(Greeks) computation as a high-value target, where quantum gradient algorithms offer simultaneous speedups across many risk factors. While Quantum Singular Value Transformation~(QSVT) has been proposed to resolve the arithmetic bottleneck in payoff loading~\cite{veronelli2025implementing}, its substantially deeper circuits render it ill-suited to NISQ hardware. We instead adopt IQAE, which retains the quadratic sampling advantage at circuit depths compatible with current simulation environments.

As the quantum hardware ecosystem matures, research has shifted toward hybrid quantum-classical architectures that pair classical pre-processing stability with quantum high-dimensional integration capabilities. Pistoia et al.~\cite{pistoia2021quantum} survey quantum algorithms for finance, identifying CCR as among the highest-priority near-term use cases, while Rebentrost et al.~\cite{rebentrost2018quantum} establish quantum speedup for portfolio optimisation---a problem sharing the same integration bottleneck. High-level synthesis platforms, most notably Classiq and Qiskit~\cite{Pathak2025}, have become essential for automating the deep, multi-step circuit generation required for CCR analytics~\cite{An2021, Auer2024}. In the current NISQ era, GPU-accelerated emulators such as NVIDIA CUDA-Q and cloud-scale state-vector simulators such as Amazon Braket SV1 provide the necessary bridge to utility, enabling controlled quantification of noise and discretization errors in a hardware-agnostic setting~\cite{Dri2025}.

To our best knowledge, this work introduces the first hybrid pipeline that replaces full distributional quantile estimation with percentile‑conditioned conditional expectation evaluated via IQAE---a formulation not yet explicitly explored in the existing quantum CCR or derivative pricing frameworks.

\section{Counterparty Credit Risk}

\subsection{Problem Definition} 

Counterparty Credit Risk on a portfolio of derivatives is measured as Potential Future Exposure (PFE). The portfolio value is computed using thousands of market scenarios over various time intervals $t$, and PFE at time $t$ is estimated as a quantile of the portfolio exposure $E(t) := \text{max}(V(\pi,t),0)$, which reflects that only positive mark-to-market values constitute credit risk to the counterparty. PFE is driven by trade cashflows, netting agreements, collateral and model parameters.

Formally, the PFE at time $t$ for a portfolio of derivatives $\pi$ under measure $M$ at a quantile $q \in (0,1)$ is given by:
\begin{align}
\text{PFE}_{M}(t,q) = Q^{q}_{M}\left[\max(V(\pi,t),\,0) \right]
\end{align}
where $Q^{q}_{M}$ denotes the $q$-th quantile operator under measure $M$.
PFE is typically estimated using Monte Carlo simulation, whereby relevant market risk factors are simulated across future time points, and the derivative mark-to-market value is evaluated on each simulated path. Portfolio exposure is subsequently aggregated at each future time point, accounting for netting and collateral. The PFE is then obtained as the relevant quantile of the resulting exposure distribution.

\subsection{Payoff Function}

In this work, we focus on an FX TARF---a series of FX forwards subject to knock-out condition on cumulative gains. The payoff usually specifies a strike price $K$, an upper barrier $B_u$, a lower barrier $B_l$ and a leverage $\alpha$. The trade typically has 13-52 weekly fixings, and on each fixing date $t$, as the underlying FX price fixes at $S_t$, the payoff is determined as follows:
\begin{align}
    p_t = 
    \begin{cases}
    S_t - K,          &\text{ if } S_t \ge B_u \\
    0,                &\text{ if } B_l < S_t < B_u \\
    \alpha(S_t - K)  &\text{ if } S_t \le B_l
    \end{cases}
\end{align}
Positive payoffs accumulate across fixing dates. The trade knocks out on the first fixing date at which the cumulative gains exceed a cap $C$. The cumulative gain is capped at $C$, and no further fixings occur.

\subsection{Potential Quantum Advantage}

Quantum advantage in CCR estimation is expected to arise from improved sampling efficiency in high quantile exposure estimation, rather than from raw computational speedup. In particular, path dependent derivatives and tail risk metrics such as PFE at the 97.5\% and 99\% confidence levels remain dominated by statistical sampling noise under classical Monte Carlo methods, once standard accuracy improvement techniques reach diminishing returns for rare but high impact scenarios \cite{glasserman2004monte,Ghamami2014}. Quantum Amplitude Estimation (QAE) directly targets this bottleneck through its quadratic improvement in sample complexity for expectation and probability estimation \cite{montanaro2015quantum,brassard2002quantum}, making it algorithmically well aligned with tail risk calculations \cite{woerner2019quantum,Egger2020}.

However, realizing this advantage is contingent on several conditions. First, the fidelity of the quantum state representation must be sufficiently high to ensure that discretization error in the stochastic dynamics is smaller than the targeted estimation error. Second, circuit depth and hardware noise must be tightly controlled, as tail risk estimation relies on resolving small differences in probability amplitudes that are highly sensitive to decoherence and gate errors. Third, quantum routines must be deployed within hybrid quantum–classical workflows, where classical systems continue to handle scenario generation, percentile selection and portfolio aggregation, while quantum subroutines are used selectively to accelerate the most computationally intensive conditional expectation steps \cite{Alcazar2022,Auer2024}.

Under these conditions, quantum advantage should be interpreted as a throughput advantage at scale, rather than near term wall clock acceleration. Our analysis suggests that this becomes plausible on fault tolerant hardware with sufficient logical qubits to support fine-grained state encoding and stable amplitude estimation over multi-step horizons.

\section{Proposed Approach}

\subsection{Quantum Adaptation}

CCR estimation for path-dependent derivatives typically requires nested Monte Carlo simulations, where outer simulations generate market trajectories and inner simulations estimate exposure distributions conditional on these trajectories~\cite{glasserman2004monte}. This results in substantial computational complexity---scaling as $\mathcal{O}(N \dot M)$---particularly when estimating high quantiles of the exposure distribution.

To enable execution on near-term quantum hardware and simulators, the original problem formulation was adapted along three dimensions. First, the temporal horizon was reduced from a full-year (52-week) exposure profile to a two-step model corresponding to two weekly intervals. This is a deliberate scoping choice that introduces an approximation of the full exposure dynamics while preserving key structural features of the TARF: the two-step model preserves the essential path-dependent and knock-out structure of the TARF---the accumulation of gains across fixing dates and the conditional termination logic---while limiting circuit depth and qubit requirements to levels tractable on current simulation infrastructure. 

Second, the stochastic dynamics of the FX rate was simplified to reduce quantum arithmetic cost. In the classical model, the relative spot evolves according to a log-normal process,
\begin{align}
S_t = S_{t-1} e^{\left(-\frac{1}{2}\sigma_w^2 + \sigma_w Z_t\right)},
\end{align}
where $Z_t \sim \mathcal{N}(0,1)$ and $\sigma_w$ denotes the weekly volatility. A direct quantum implementation of this update would require repeated multiplications and comparisons, leading to substantial circuit width. To reduce the arithmetic cost, we introduce a rescaled variable $y$ through:
\begin{align}
x = 1 - By,
\end{align}
where $x$ denotes the relative spot normalized by the initial spot $S_0$, and $B$ is a scaling factor chosen so that the discretized values of $y$ fit within the available register size. For clarity, we use $y$ for the continuous rescaled variable and $\hat{y}$ for its discretized representation throughout.

Neglecting terms of order $B^2$, products of spot factors can be subsequently approximated by sums in the rescaled variables. For two weekly steps beyond the initial spot,
\begin{align}
x_0 &= 1 - By_0,\\
x_1 &= (1 - By_0)(1 - By_1) \approx 1 - B(y_0 + y_1).
\end{align}
This approximation eliminates costly quantum multiplications, making the payoff arithmetic largely additive and substantially reducing the circuit width. The approximation is valid when $B\sigma_w \ll 1$: for weekly FX volatility $\sigma_w \approx 0.01$--$0.03$ and $B$ chosen such that $B\sigma_w < 0.05$, the $\mathcal{O}(B^2)$ residual contributes less
than $0.1\%$ error to the spot product---well within the IQAE
precision target of $\epsilon = 0.05$.

Finally, the continuous state space of FX rates was discretized using a finite number of qubits. Experiments were conducted with 4, 5, and 6 qubits to represent the underlying asset, yielding $2^n$ discrete bins for an $n$-qubit configuration. The largest configuration (6~qubits, 64~bins) in circuits exceeding 30 qubits when including ancillary registers required for arithmetic and amplitude encoding.

The resulting model captures a reduced yet representative instance of CCR estimation for a path-dependent FX derivative (JPY/USD), enabling a controlled comparison between classical and quantum approaches under realistic hardware constraints.
This reduction also enables a hybrid workflow in which percentile selection is performed classically, while exposure estimation is carried out using quantum amplitude estimation.

In the implemented workflow, the two-step horizon is realized using a hybrid construction. The first time step is evaluated classically by computing the percentile value of the 1-week FX distribution, mapping it to the rescaled variable $y_0$, and discretizing it into $\hat{y}_0$. The second time step is modeled quantumly, where the evolution and payoff are computed using amplitude estimation.

This decomposition is justifiable under the assumption that the continuation exposure is near monotone in the 1-week FX state. In the ideal monotone case, if the exposure after one week can be written as \(E_1 = g(X_1)\), where \(X_1\) is the 1-week FX state and \(g\) is monotone, then the \(q\)-quantile of exposure satisfies \(Q_q(E_1)=g(Q_q(X_1))\). In this case, selecting the percentile value of the first time step classically and estimating the conditional continuation exposure quantumly yields the corresponding tail-exposure estimate. In non-monotone settings, this approximation may introduce bias relative to full quantile estimation. Consequently, the method estimates conditional exposure at a percentile-conditioned state rather than computing the quantile of the full exposure distribution.

This formulation reduces the quantum resource requirements by shifting part of the stochastic evolution to classical preprocessing, while preserving the structure of the exposure calculation.

\subsection{Quantum and Classical Complexity}

Classical CCR estimation for path-dependent derivatives often relies on nested Monte Carlo methods. Let $N$ be the number of outer scenarios and $M$ the number of inner simulations per scenario. The total computational cost scales as $\mathcal{O}(N \cdot M)$, with additional overhead for quantile estimation~\cite{glasserman2004monte}.

Quantum Amplitude Estimation (QAE) provides a quadratic improvement in the estimation of expected values and probabilities, reducing the sample complexity from $\mathcal{O}(1/\epsilon^2)$ to $\mathcal{O}(1/\epsilon)$ for a target precision $\epsilon$~\cite{montanaro2015quantum, brassard2002quantum}. This advantage extends to financial risk metrics that can be expressed in terms of expectation or probability estimation~\cite{woerner2019quantum}. In the present work, this advantage is leveraged within a hybrid workflow, where exposure is estimated quantumly for classically selected percentile inputs.

In this work, the CCR problem is mapped to a quantum operator $\mathcal{A}$ that prepares a quantum state encoding the payoff distribution across all simulated paths and time steps. The probability amplitude of a designated measurement qubit corresponds to the quantity of interest, which is subsequently estimated using amplitude estimation techniques.

Since canonical QAE relies on Quantum Phase Estimation and deep circuit constructions, we employ Iterative Quantum Amplitude Estimation (IQAE), which retains the quadratic speedup while avoiding the need for quantum Fourier transforms and controlled powers of the Grover operator~\cite{grinko2021iterative}. This makes it suitable for noisy intermediate-scale quantum (NISQ) devices.

\subsection{Quantum Approach}

The quantum algorithm implements a coherent version of the exposure computation, combining probabilistic state preparation, arithmetic evaluation of the payoff, and amplitude estimation.

\textbf{State Preparation.}  
Quantum registers are initialized using a piecewise-polynomial state preparation circuit to encode the discretized log-normal distribution over a finite grid, resulting in a superposition
\begin{align}
\sum_z \sqrt{p(z)} \ket{z},
\end{align}
where $z$ indexes the discretized realizations and $p(z)$ denotes their corresponding probabilities.

\textbf{Path Evolution and Payoff Computation.} 
For each basis state $\ket{z}$, the FX evolution and payoff are computed using reversible arithmetic. Following the linearized formulation described in Section~4.1, the spot evolution is expressed in terms of additive contributions of the rescaled variables $y_i$. In the implementation, the input variable $\hat{y}_0$ corresponds to a discretized representation of the rescaled variable $y_0$ associated with the first (classically evaluated) time step. The payoff is accumulated in a dedicated register (denoted \texttt{asset}) by iterating over time steps and applying the contract-specific payoff logic.

In particular, the implementation:
\begin{itemize}
    \item stores intermediate values of the rescaled variables in dedicated registers,
    \item accumulates the payoff across time steps using an auxiliary register,
    \item incorporates conditional logic corresponding to the path-dependent structure of the derivative.
\end{itemize}

\textbf{Amplitude Encoding of the Payoff.}  
Since the accumulated payoff may take both positive and negative values, it is first mapped to the unit interval via an affine transformation
\begin{align}
g(a) = \frac{-(a + 1 - 2^{(\mathrm{asset\_size}-1)})}{2^{\mathrm{asset\_size}} - 1}
\end{align}

A controlled rotation is then applied such that the amplitude of a designated indicator qubit encodes the square root of the rescaled payoff:

{
\footnotesize
\begin{align}
\mathcal{A}\ket{0} = \sum_z \sqrt{p(z)} \ket{z} & \left( \sqrt{1-g(a(z))}\ket{0} + \sqrt{g(a(z))}\ket{1} \right)
\end{align}
}
Thus, the probability of measuring the indicator qubit in the $\ket{1}$ state corresponds to the expectation value of the transformed payoff $g(a)$, from which the exposure in the original scale can be recovered via classical post-processing.

\textbf{Amplitude Estimation}. The operator $\mathcal{A}$ is used within an Iterative Quantum Amplitude Estimation (IQAE) routine to estimate the amplitude associated with the indicator qubit. IQAE iteratively refines the estimate while avoiding the use of quantum phase estimation, making it suitable for circuits of moderate depth. The estimated amplitude is then classically post-processed to recover the expected exposure corresponding to a given input state.

\begin{algorithm}[!t]
\caption{Quantum Monte Carlo for Exposure Estimation}
\begin{algorithmic}[1]

\STATE \textbf{Input:} Discretized distribution $p(z)$, payoff function, $T=2$
\STATE Compute the classical one-step percentile value and map it to $y_0$, then discretize to $\hat{y}_0$
\STATE Prepare quantum state
\[
\ket{\psi} = \sum_z \sqrt{p(z)} \ket{z}
\]
\STATE Encode the stochastic evolution of the second time step in this state
\STATE Use $\hat{y}_0$ as input to the payoff computation
\STATE Map payoff: $a \rightarrow g(a) \in [0,1]$
\STATE Apply controlled rotation:
\[
\ket{z}\ket{0} \rightarrow \ket{z} \left( \sqrt{1-g(a(z))}\ket{0} + \sqrt{g(a(z))}\ket{1} \right)
\]
\STATE Apply IQAE to estimate the expected payoff (exposure) corresponding to the input state
\STATE \textbf{Output:} Estimated exposure for the selected percentile input
\end{algorithmic}
\end{algorithm}

\begin{figure*}[!t]
\centering
\includegraphics[width=\textwidth]{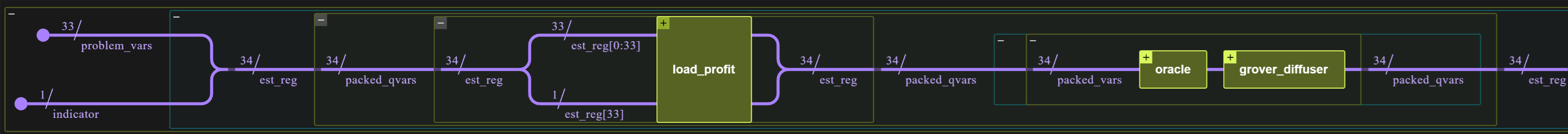} 
\caption{Circuit diagram of the proposed Quantum Monte Carlo solution synthesized on the Classiq platform.}
\label{fig:classiq_circuit}
\end{figure*}

High-quantile exposure levels are selected classically. For a given percentile level $P_0$, a corresponding 1-step FX value is computed from the log-normal distribution and mapped to the rescaled variable $y_0$, which is discretized into $\hat{y}_0$. The quantum algorithm then models the subsequent time step and estimates the expected gain or loss conditional on this input. This hybrid construction enables a two-step exposure calculation while requiring quantum resources for only a single stochastic evolution step.

The IQAE routine is configured with precision parameter $\epsilon = 0.05$ and confidence parameter $\alpha = 0.01$, corresponding to a confidence level of $1-\alpha = 99\%$. These parameters introduce a controlled approximation error in the amplitude estimation, which contributes to the overall uncertainty of the reported exposure values.

\textbf{Algorithm Overview.} An overview of the proposed Quantum Monte Carlo algorithm for exposure estimation is given in Algorithm 1.

\subsection{Circuit Synthesis}

Circuit synthesis was performed using the Classiq platform, with an optimisation objective to minimize the circuit width, that is, the number of qubits. Fig. \ref{fig:classiq_circuit} shows a block diagram of the synthesized circuit. The pipeline is decomposed into two primary unitary primitives: the state preparation operator $\mathcal{A}$ and the Grover amplification operator $\mathcal{Q}$. The \texttt{load\_profit} block serves as the operator $\mathcal{A}$, which discretizes the underlying stochastic FX process and encodes the non-linear, path-dependent payoff logic---incorporating cumulative gains and knock-out boundaries---into the probability amplitude of a designated objective qubit. Subsequently, the \texttt{phase oracle} and \texttt{grover diffuser} blocks collectively define the $\mathcal{Q}$ operator; the oracle applies a phase flip to the ``good'' states marked by the objective qubit, while the diffuser performs a reflection about the initial state. As delineated in \textit{Algorithm 1}, this circuit is executed iteratively within a hybrid quantum-classical loop.


\section{Results and Discussions}

\subsection{Experiment Scope}

The experimental evaluation was designed to assess correctness and feasibility under realistic hardware constraints. All experiments were conducted under the following conditions:
\begin{itemize}
    \item Two-step FX model corresponding to a two-week horizon
    \item Discretized state space using 4--6 qubits
    \item Estimation of high-quantile exposure, with reported results at the 97.5th and 99th percentiles
    \item Comparison against classical Monte Carlo benchmarks
\end{itemize}

Circuit metrics, including qubit count and circuit depth, were recorded to assess resource requirements as the discretization resolution increased. In particular, increasing the number of qubits used to represent the FX distribution (from 4 to 6) led to a measurable growth in circuit size and simulation cost, reflecting the trade-off between improved discretization accuracy and computational feasibility.

Simulations were executed on:
\begin{itemize}
    \item Classiq simulator for circuit synthesis and validation
    \item Amazon Braket SV1 simulator for larger-scale circuit execution (6-qubit configurations)
    \item NVIDIA CUDA-Q for high-performance emulation
\end{itemize}

The same quantum circuits were executed using NVIDIA CUDA-Q for both 5- and 6-qubit configurations. The resulting exposure estimates were consistent with those obtained from the Classiq simulator and Braket SV1. However, CUDA-Q provided a substantial reduction in execution time. 

For the 6-qubit configuration, a representative simulation that required approximately 142 minutes on Amazon Braket SV1 completed in approximately 67 seconds on a GPU-enabled system using CUDA-Q, corresponding to a speedup of nearly 30$\times$. This highlights the major computational cost associated with high-fidelity quantum circuit simulation, and the importance of leveraging accelerated classical infrastructure, such as GPU-based simulators and High-Performance Computing (HPC) environments, for prototyping and validation.

While such accelerations are effective for small- to mid-scale problems, they do not change the underlying exponential scaling of classical simulation. The anticipated advantage of quantum approaches ultimately relies on executing larger instances directly on quantum hardware as it matures.

\subsection{Main Results}

\subsubsection{Validation against Classical Monte Carlo}

The quantum results were benchmarked against classical nested MC simulations across eight representative FX TARF Trades (EX1--EX8), evaluating PFE at the 97.5th and 99th percentiles. The trade and market data details are provided in Table~\ref{tab:trade_data}. The reference FX spot price is \$148, $\alpha$=2 and trade expiry is 2 weeks for all trades.

Overall, the quantum estimates demonstrate strong agreement with classical estimates. Fig.~\ref{fig:scatter} shows that the relative errors are typically constrained between 1\% and 8\%, excluding the low-exposure outlier discussed below. This level of agreement is notable given the simplified two-step model, the discretization of the state space, and the finite precision and statistical uncertainty inherent in the quantum amplitude estimation procedure. 

\begin{table}[!t]
\centering
\caption{Trade and Market Data}
\label{tab:trade_data}
\begin{tabular}{lccccc}
\hline
\textbf{Case} & \textbf{Strike} & \textbf{Bl} & \textbf{Bu} & \textbf{Target} & \textbf{Volume} \\ \hline
EX1  & 150 & 150 & 151 & 4  & 12\% \\
EX2  & 150 & 150 & 152 & 6  & 14\% \\
EX3  & 151 & 151 & 154 & 8  & 16\% \\
EX4  & 151 & 151 & 154 & 10 & 20\% \\
EX5  & 151 & 151 & 154 & 12 & 24\% \\
EX6  & 153 & 153 & 154 & 4  & 12\% \\
EX7  & 154 & 154 & 156 & 6  & 14\% \\
EX8  & 148 & 148 & 150 & 6  & 14\% \\ \hline
\end{tabular}
\end{table}



\definecolor{cb_blue}{RGB}{0,114,178}
\definecolor{cb_vermillion}{RGB}{213,94,0}

\begin{figure}[!t]
\centering
\begin{tikzpicture}
\begin{groupplot}[
    group style={group size=1 by 2}, 
    width=1.0\columnwidth,
    grid=both,
    minor tick num=1,
    xlabel={Classical PFE},
    ylabel={Quantum PFE},
    legend style={font=\small, draw=none, at={(0.03,0.97)}, anchor=north west},
]

\nextgroupplot[
    title={97.5\% percentile},
    xmin=0, xmax=16,
    ymin=0, ymax=16,
]
\addplot[domain=0:16, samples=2, black, dashed] {x};
\addlegendentry{$y=x$}

\addplot[only marks, color=cb_blue, mark=*] coordinates {
    (5.84,6.6430)
    (7.43,7.5160)
    (6.85,6.9300)
    (10.28,10.7307)
    (13.77,15.0250)
    (0.87,1.0123)
    (0.55,0.9793)
    (11.35,11.8147)
};
\addlegendentry{\#Q = 5}

\addplot[only marks, color=cb_vermillion, mark=square, thick] coordinates {
    (5.84,5.7893)
    (7.43,7.3070)
    (6.85,6.5163)
    (10.28,10.2630)
    (13.77,14.2090)
    (0.87,0.8835)
    (0.55,0.6703)
    (11.35,11.4700)
};
\addlegendentry{\#Q = 6}

\nextgroupplot[
    title={99\% percentile},
    xmin=0, xmax=20,
    ymin=0, ymax=20,
    legend style={font=\small, draw=none, at={(0.03,0.97)}, anchor=north west},
]
\addplot[domain=0:20, samples=2, black, dashed] {x};
\addlegendentry{$y=x$}

\addplot[only marks, color=cb_blue, mark=*] coordinates {
    (7.69,8.5807)
    (9.61,9.7673)
    (9.33,9.6593)
    (13.38,13.8597)
    (17.67,18.9080)
    (2.23,2.2590)
    (2.16,2.5050)
    (13.58,14.1473)
};
\addlegendentry{\#Q = 5}

\end{groupplot}
\end{tikzpicture}
\caption{Comparison between classical and quantum PFE estimates across all scenarios. The 97.5\% panel includes both 5-qubit and 6-qubit (SV1) results, while the 99\% panel includes only 5-qubit results.}
\label{fig:scatter}
\end{figure}
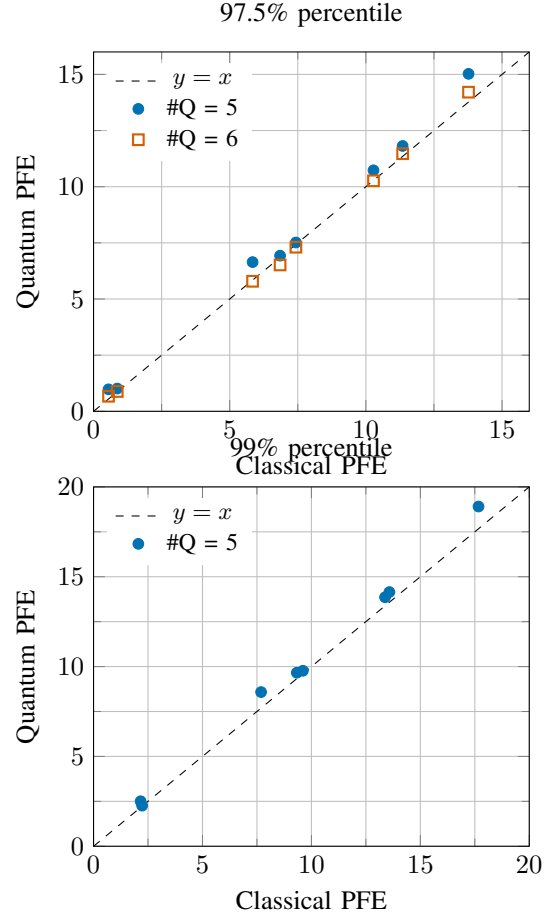

\subsubsection{Results on Braket SV1}

To assess the impact of increased state resolution, additional experiments were conducted using a 6-qubit discretization of the FX variable on the Amazon Braket SV1 simulator. Due to the significantly higher computational cost of these simulations, only the 97.5\% percentile was evaluated, and each scenario was executed a limited number of times (between 2 and 5 repetitions).

Across the scenarios evaluated at the 97.5\% level, the 6-qubit results remain consistent with the trends observed in the 4- and 5-qubit experiments and exhibit good agreement with the classical baseline, with deviations primarily attributable to discretization and statistical uncertainty.


\definecolor{cb_blue}{RGB}{0,114,178}
\definecolor{cb_vermillion}{RGB}{213,94,0}
\definecolor{cb_green}{RGB}{0,158,115}

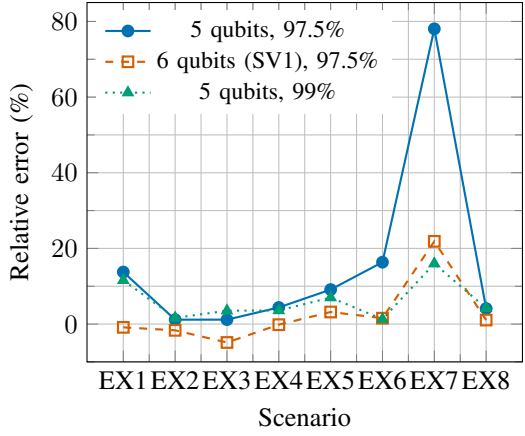
\begin{figure}[!t]
\centering
\begin{tikzpicture}
\begin{axis}[
    width=1.0\columnwidth,
    grid=both,
    minor tick num=1,
    xlabel={Scenario},
    ylabel={Relative error (\%)},
    symbolic x coords={EX1,EX2,EX3,EX4,EX5,EX6,EX7,EX8},
    xtick=data,
    ymin=-10, ymax=85,
    legend style={font=\small, draw=none, at={(0.02,0.98)}, anchor=north west},
]
\addplot[color=cb_blue, mark=*, thick, solid] coordinates {
    (EX1,13.7500)
    (EX2,1.1575)
    (EX3,1.1679)
    (EX4,4.3839)
    (EX5,9.1140)
    (EX6,16.3602)
    (EX7,78.0485)
    (EX8,4.0940)
};
\addlegendentry{5 qubits, 97.5\%}

\addplot[color=cb_vermillion, mark=square, thick, dashed, mark options={solid}] coordinates {
    (EX1,-0.8676)
    (EX2,-1.6555)
    (EX3,-4.8723)
    (EX4,-0.1654)
    (EX5,3.1881)
    (EX6,1.5517)
    (EX7,21.8636)
    (EX8,1.0573)
};
\addlegendentry{6 qubits (SV1), 97.5\%}

\addplot[color=cb_green, mark=triangle*, thick, dotted, mark options={solid}] coordinates {
    (EX1,11.5821)
    (EX2,1.6372)
    (EX3,3.5298)
    (EX4,3.5850)
    (EX5,7.0062)
    (EX6,1.3004)
    (EX7,15.9722)
    (EX8,4.1777)
};
\addlegendentry{5 qubits, 99\%}
\end{axis}
\end{tikzpicture}
\caption{Relative error across scenarios. The 6-qubit SV1 results at the 97.5\% percentile generally reduce the discretization-induced bias observed in the 5-qubit configuration, although the magnitude of improvement varies by scenario.}
\label{fig:error_plot}
\end{figure}

Increasing the number of qubits leads to a finer discretization of the underlying distribution, which reduces quantization effects and improves the resolution of percentile-based inputs, although the observed improvement is moderated by sampling variability in the current experiments.

Quantitatively, the 6-qubit results remain closely aligned with the classical baseline across all scenarios. For example, in scenario EX1, the classical 97.5\% exposure of 5.84 is matched by quantum estimates in the range 5.60–5.90. Similarly, in EX5, the classical value of 13.77 is matched by quantum estimates in the range 13.63–14.95.

Across scenarios, the spread of repeated quantum estimates reflects both discretization effects and the limited number of samples, but does not exhibit systematic degradation relative to the lower-qubit configurations. In several cases, the 6-qubit discretization slightly reduces bias relative to the classical value, although this improvement is not uniform due to sampling variability.

\subsubsection{Systematic Bias and Error Sources}

Across the scenarios considered, the quantum estimates consistently exceed the corresponding classical values, indicating a systematic positive bias. This effect can be attributed to the discretization of the FX variable and transformation of the classical percentile value into the integer representation $\hat{y}_0$. As this transformation involves rounding, it imposes a bias that propagates through the payoff computation.

Discretization also induces a finite binning of the underlying distribution. For the 5-qubit implementation, the FX state space is represented using $2^5 = 32$ discrete bins. In several scenarios, both the 97.5\% and 99\% percentiles fall within the same discretization bin. As a result, the quantum computation effectively evaluates exposure at the same discretized input for both percentiles, leading to similar or correlated errors across quantile levels.

Statistical uncertainty arising from IQAE also contributes to variability in the results, although this effect appears secondary compared to discretization. The structure of the relative error across scenarios is illustrated in Fig.~\ref{fig:error_plot}.

\subsubsection{Outlier Behavior}

One scenario (EX7) exhibits significantly larger deviation, with an error of approximately 80\% at the 97.5th percentile. This behavior is associated with a low-exposure regime in which discretization effects dominate the signal. In such cases, small absolute differences in the discretized input can lead to large relative errors in the resulting exposure.

At the 99th percentile, the error for this scenario is reduced to approximately 16\%, further supporting the interpretation that discretization sensitivity is amplified in lower-exposure regions.

\subsubsection{Effect of Percentile Level}

Estimates at the 99th percentile are shown to be more stable and exhibit lower relative error than at the 97.5th percentile,. This can be attributed to smoother exposure behavior in the extreme tail and lower sensitivity to discretization at high values.

\subsubsection{Discretization-Induced Quantization Effects}

The finite resolution of the quantum representation introduces a quantization effect in the evaluation of percentile-based exposure. In the present implementation, the discretized FX variable is encoded using 4 to 6 qubits, resulting in a progressively finer partitioning of the state space as the number of qubits increases.

As a consequence, nearby percentile values may map to the same discretized state. For the 5-qubit configuration, the 97.5\% and 99\% percentiles often fall into the same bin of the discretized distribution. This leads to similar input values for the quantum computation and reduces the effective distinction between these percentiles at the quantum level.

This effect explains both the similarity in the quantum estimates across percentile levels and the observed structure of the errors. Increasing the number of qubits, as demonstrated in the 6-qubit experiments, refines the discretization, allowing different percentiles to be resolved more accurately and reducing this source of bias.

\subsection{Resource Scaling Analysis}

To assess the scalability of the proposed framework beyond the reduced two-step demonstration, we synthesized extended versions of the forward payoff-evaluation oracle with increasing temporal horizons while preserving the same TARF payoff logic. The analysis focuses on the scaling of the synthesized oracle and an IQAE-compatible Grover-iterate proxy, rather than a complete end-to-end production PFE workflow.

Table~\ref{tab:resource_scaling_weeks} reports the synthesized resources for the forward oracle $A$ as the number of weekly fixing steps increases from 5 to 52. The circuit width scales approximately linearly with the number of weeks, reaching 306 qubits for the 52-week configuration. Circuit depth and CX count grow more rapidly, reflecting the repeated reversible arithmetic, comparison logic, payoff accumulation, and uncomputation required by the path-dependent structure.

\begin{table}[htbp]
\centering
\caption{Resource scaling of the synthesized forward payoff-evaluation oracle $A$ as the number of weekly fixing steps increases.}
\label{tab:resource_scaling_weeks}
\begin{tabular}{c|r|r|r}
\hline
Weeks & Width & Depth & CX Count \\
\hline
5  & 50  & 14,381  & 12,970 \\
10 & 82  & 52,520  & 51,152 \\
15 & 109 & 89,147  & 85,652 \\
20 & 138 & 144,036 & 143,784 \\
25 & 165 & 195,549 & 198,966 \\
30 & 190 & 229,137 & 225,656 \\
35 & 219 & 312,849 & 331,090 \\
40 & 244 & 358,644 & 379,600 \\
52 & 306 & 516,389 & 552,188 \\
\hline
\end{tabular}
\end{table}

\begin{figure}[htbp]
\centering
\begin{tikzpicture}
\begin{groupplot}[
    group style={group size=2 by 1, horizontal sep=0.8cm}, 
    width=0.6\columnwidth,  
    height=0.6\columnwidth,  
    grid=both,
    xlabel={Weeks},
    title style={font=\normalsize}, 
    xlabel style={font=\small},     
    tick label style={font=\small}, 
]

\nextgroupplot[
    title={Width},
    ymin=0,
    ymax=330,
]
\addplot[
    blue,
    mark=*,
    thick
] coordinates {
    (5,50)
    (10,82)
    (15,109)
    (20,138)
    (25,165)
    (30,190)
    (35,219)
    (40,244)
    (52,306)
};

\nextgroupplot[
    title={Depth (in millions)},
    ymin=0,
    ymax=3.2,
]
\addplot[
    red,
    mark=*,
    thick
] coordinates {
    (5,0.014381)
    (10,0.052520)
    (15,0.089147)
    (20,0.144036)
    (25,0.195549)
    (30,0.229137)
    (35,0.312849)
    (40,0.358644)
    (52,0.516389)
};

\addplot[
    orange,
    mark=square*,
    thick
] coordinates {
    (5,0.076650)
    (10,0.254444)
    (15,0.443136)
    (20,0.718368)
    (25,1.000186)
    (30,1.240317)
    (35,1.680930)
    (40,2.000813)
    (52,3.009277)
};

\end{groupplot}
\end{tikzpicture}
\caption{Circuit width and depth scaling with weekly fixing steps. Depth is shown in millions of circuit layers. In the depth panel, the lower curve corresponds to the forward payoff-evaluation oracle $A$, while the upper curve corresponds to the Grover-style iterate proxy $G$ used to estimate IQAE overhead.}
\label{fig:resource_scaling_weeks}
\end{figure}
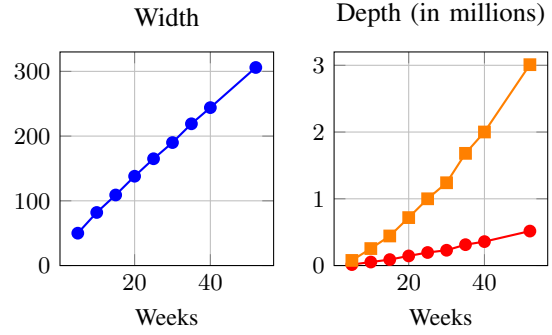

To estimate the overhead introduced by IQAE, we additionally synthesized a Grover-style iterate $G$ representative of the repeated operations used in amplitude estimation. As shown in Fig.~\ref{fig:resource_scaling_weeks}, this Grover-iterate proxy increases the circuit depth by approximately a factor of 5--6 relative to the forward oracle, while leaving the qubit count essentially unchanged. For the 52-week configuration, the Grover proxy reaches depth approximately $3.0\times10^6$ and CX count approximately $2.4\times10^6$.

A complete IQAE execution requires repeated applications of the Grover iterate. For the parameters used in this work, $\epsilon=0.05$ and $\alpha=0.01$, the effective number of Grover-oracle applications is expected to be on the order of several tens. Using a representative value of approximately 30 applications, the 52-week configuration corresponds to an estimated logical execution depth on the order of $10^8$ gate layers and a total two-qubit gate count also on the order of $10^8$.

Taken together, these estimates provide a first-order indication of the quantum resources required for long-horizon path-dependent exposure estimation. For a 52-week horizon, the synthesized workload requires on the order of 300 logical qubits, Grover-iterate depths of order $10^6$, and full IQAE execution depths approaching $10^8$ logical gate layers. This places the workload beyond current NISQ devices, but within a resource regime relevant for future fault-tolerant quantum processors, where sufficiently low logical error rates and long coherent execution windows would be required to realize practical utility for PFE estimation.

\section{Conclusion and Future Directions}

We have demonstrated that IQAE-based estimation is able to approximate the high-quantile exposure for path-dependent FX derivatives with $1\%$--$8\%$ relative error against classical benchmarks, within a hybrid quantum--classical framework that preserves the essential knock-out and path-dependent structure of the TARF under a reduced two-step formulation. However, this approach estimates exposure via conditional expectation and does not recover the full distributional quantile underlying PFE. Quantitative scaling analysis indicates
$300$ logical qubits would be required to enable full-horizon (52-week) exposure estimation; however, the associated execution depth of the order of $10^8$ gate layers also implies substantial runtime and error-correction for practical execution. 

Our central insight is that the dominant constraint on accuracy arises not from the amplitude estimation procedure---which performs in line with theoretical predictions---but the fidelity of classical-to-quantum state encoding. This reframes the path to quantum advantage in CCR estimation under NISQ constraints; the key bottleneck lies in quantum state preparation rather than further refinement of amplitude estimation routines. Future work should therefore target three fronts:

\begin{itemize}
    \item \textbf{State encoding:} Higher-fidelity representation
    of continuous stochastic processes onto finite qubit registers.

    \item \textbf{Richer dynamics:} Extension to stochastic
    volatility, regime changes, and jump processes for more
    realistic derivative pricing.

    \item \textbf{Hardware validation:} Empirical execution on
    physical quantum devices to close the simulation-to-device
    performance gap and establish practical conditions for quantum
    utility.
\end{itemize}

\bibliographystyle{ieeetr} 
\bibliography{references} 

\end{document}